\documentclass[12pt,a4]{amsart}

\usepackage[utf8]{inputenc}
\usepackage[T1]{fontenc}
\usepackage{amsmath,comment,color,float,graphicx}
\usepackage{amssymb}
\usepackage{amsthm}
 \usepackage[numbers]{natbib}

\usepackage{graphicx}
\usepackage{enumerate}
\usepackage{bm}
\usepackage{setspace} 

\usepackage{natbib} 
\usepackage{url} 

\usepackage[plain,noend,ruled,linesnumbered]{algorithm2e}
\usepackage[colorlinks=true,allcolors=blue]{hyperref}

\newsavebox{\algbox}

\newcommand{\bfQ}{{ \bf Q} }
\newcommand{\bfu}{{\bf u}}

\newcommand{\bfTh}{{\bf \Theta} }

\newcommand{\bfx}{{\bf x}}

\newcommand{\bfZ}{{\bf Z}}

\newcommand{\bfz}{{\bf z}}
\newcommand{\bfS}{{\bf S}}

\newcommand{\bfX}{{\bf X}}

\newcommand{\bre}{\begin{remark}\em }
\newcommand{\ere}{\end{remark}}
\def\1{\ensuremath{\mathrm{1}\hspace{-.35em} \mathrm{1}}} 

\def\E{{\mathbb E}}

\def\P{\mathbb{P}}

\newtheorem{lemma}{Lemma}[section]

\newtheorem{theorem}[lemma]{Theorem}

\newtheorem{definition}[lemma]{Definition}

\newtheorem{remark}[lemma]{Remark}

\begin{document}

\date{}
\today
\title[A stable sums approach to infer high return levels]{
Stable sums to infer high return levels of  multivariate rainfall time series
}

\author[G. Buritic\'a]{Gloria Buritic\'a
}
\address{Laboratoire de Probabilit\'es, Statistiques et Mod\'elisation\\
Sorbonne Universit\'e,
75005, Paris, France}
\email{gloria.buritica@sorbonne-universite.fr}

\author[P. Naveau]{Philippe Naveau}
\address{Laboratoire de Sciences du Climat et de l'Environnement, EstimR\\ IPSL-CNRS,
91191, Gif-sur-Yvette, France }
\email{philippe.naveau@lsce.ipsl.fr
}

\thanks{
    The authors gratefully acknowledge John Nolan for kindly providing us with the Stable software and Olivier Wintenberger for discussions on the topic.
Part of this  work was supported by the  DAMOCLES-COST-ACTION on compound events, the French national program (FRAISE-LEFE/INSU
 and  80 PRIME CNRS-INSU), and the European H2020 XAIDA (Grant agreement ID: 101003469). 
 The authors also acknowledge the support of the French Agence Nationale de la Recherche (ANR) under reference ANR-20-CE40-0025-01 (T-REX project), 
 and the ANR-Melody. }

\maketitle

\begin{abstract}
Heavy rainfall distributional modeling is essential in any impact studies linked to the water cycle, e.g.\ flood risks. Still, statistical analyses that both take into account the temporal and multivariate nature of extreme rainfall are rare, and often,  a complex de-clustering step is needed to make extreme rainfall temporally independent. A natural question is how to bypass this de-clustering in a multivariate context. To address this issue, we introduce the stable sums method. Our goal is to incorporate time and space extreme dependencies in the analysis of heavy tails. To reach our goal, we build on large deviations of stationary regularly varying time series. Numerical experiments demonstrate that our novel approach enhances return levels inference in two ways. First, it is robust concerning time dependencies. We implement it alike on independent and dependent observations. In the univariate setting, it improves the accuracy of confidence intervals compared to the main estimators requiring temporal de-clustering. Second, it thoughtfully integrates the spatial dependencies. In simulation, the multivariate stable sums method has a smaller mean squared error than its component-wise implementation. We apply our method to infer high return levels of daily fall precipitation amounts from a national network of weather stations in France. \\
   \noindent
   
   \noindent
   \hspace{15pt}
    {\Small	
    }  
\end{abstract}

\setstretch{1.7} 

\section{Introduction}\label{sec:introduction}

Nowadays, extreme value theory \cite{coles:bawa:trenner:dorazio:2001} is frequently    applied  to  meteorological   time series   to capture  extremal 
climatological features in   temperatures, winds, precipitation and other atmospheric  variables
\cite[see, e.g.][]{kharin2013,Toulemonde13,Zscheischler20}. 
For example, due to its high societal impacts in terms of flooding, heavy rainfall have been analyzed at various spatial and temporal scales
\cite[see, e.g.][]{ipcc2021online}.  
In particular,  
storms/fronts  duration and  spatial coverage can produce potential  temporal and spatial dependencies  among recordings from  nearby weather stations
\cite[see, e.g.][]{huser2020}. 
In this multivariate context, the analysis of consecutive extremes, even in the stationary case, can be theoretically complex 
\cite[see, e.g.][]{buritica:mikosch:wintenberger:2021,basrak:planinic:soulier:2018}.  
Although  marginal behaviors of heavy rainfall is today well modeled, the temporal dynamic is rarely taken in account in applied studies, especially for multivariate time series
\cite[see, e.g.][]{Evin2017,asa18,cooley12,fawcet:walshaw:2014}.
To produce accurate high return level estimates,
we propose a novel approach to jointly incorporate the  temporal  dependence and the multidimensional structure  among heavy rainfall. 
This joint  modeling appears necessary to perform  a full   risk assessment, as ignored correlations may lead  to erroneous  confidence intervals.
The latter is particularly important when the practitioner has to provide them about extreme occurrences, i.e.\ extrapolating beyond the largest observed value. 
\par 
The practical goal of our study is to infer the 50 years return levels of fall daily rainfall from a network of weather station in France, while taking in account the multivariate dependence and the temporal memories. {The theoretical added value of our approach is that we address the extremal multivariate structure without assuming temporal independence. Moreover, our methodology does not require to decluster the $d$-dimensional time series to make observations independent in the upper tail. Declustering is particularly challenging in a multivariate context \cite{robert:2008}.} In many cases, for precise high return levels inference, declustering is unavoidable when implementing the Pareto-based methods like {\it block maxima} and {\it peaks over thresholds}  \cite[see e.g.][]{coles:bawa:trenner:dorazio:2001,chavez:davison:2012}.
To bypass these hurdles, 
we build on a stable sum method.
This new approach  takes its roots  in  large deviation principles of sums and central limit theory for weakly dependent stationary regularly varying time series in \cite{buritica:meyer:mikosch:wintenberger:2021}.
\par
We explain our stable sums method in Section~\ref{sub:sec:asymptotics}. Concerning its implementation, Section \ref{sec:stable:sums} details the ingredients of our algorithm and its assumptions. 
The important step of setting the inputs of our algorithm is treated there.
Our simulation study is described in Section \ref{sec:SS}. 
Univariate and multivariate models are investigated.
Comparisons with the main practical approaches in extreme value theory requiring declustering are implemented and commented. 
In Section \ref{sec:case:study}, we analize in depth a France rainfall dataset.
 The theoretical aspects of our method are deferred to Section \ref{sec:Asymptotic:theory}. 
Section~\ref{sec:conclusion} discusses future perspectives.


\subsection{Motivation}\label{sub:sec:motivation}
\begin{figure}[htp]
\centering
\includegraphics[height=0.8\textwidth,width=1\textwidth]{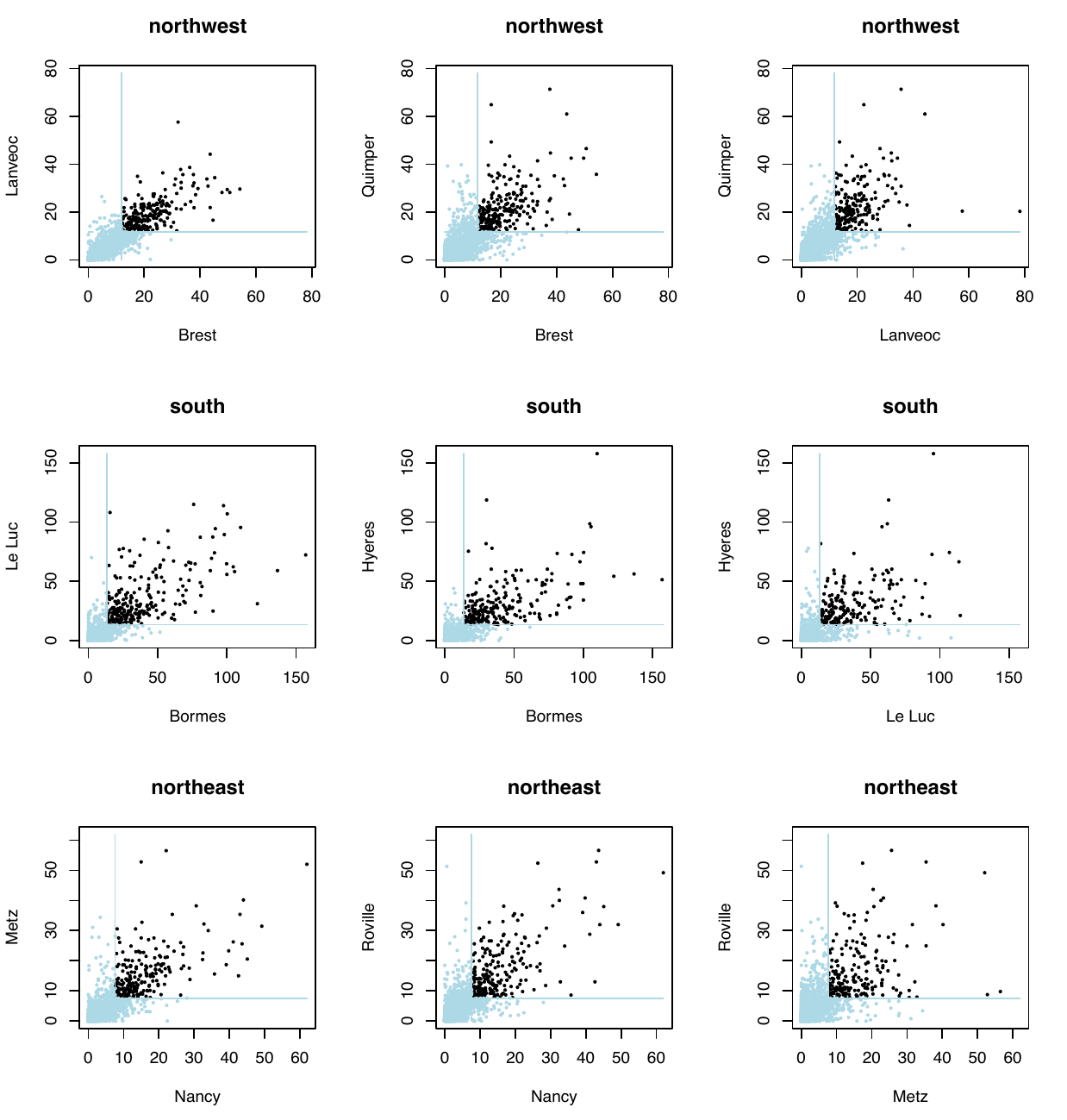}
   \caption{Scatter plots of fall daily rainfall in France from 1976 to 2015.
   The top, middle, and bottom panels refer to three climatological regions: continental (northwest), oceanic (west), and mediterranean (south), respectively.
   Simultaneous exceedances of the $95$-th order statistic of the daily maxima of a region 
   are in black.}
   \label{fig:scatter_plot}
\end{figure}

For our case study, we analyze daily precipitation from a national network in France from 1976 to 2015. To contrast different climates, we choose three stations in three different regions:  oceanic in the northwest  (Brest, Lanveoc, and Quimper),  mediterranean in the south (Hyeres, Bormes-les-Mimosas and Le Luc), and continental in the northeast (Metz, Nancy, and Roville). 
Concerning seasonality, we will focus on  Fall (September, October, and November) as heavy rainfall has been the strongest in France during this season. 
Concerning marginal behaviors, records within the same region reach similar precipitation intensity levels. For example, the south of France registers higher precipitation amounts than the other two regions, but the south attains high levels at a similar rate; see Figure~\ref{fig:scatter_plot}. 
While it is reasonable to assume independence between regions, the stations' spatial proximity within a region imposes a tri-variate analysis by region. Figure~\ref{fig:scatter_plot} illustrates how high rainfall values often co-occur at two close stations pointing to a spatial dependence of large values. 
We assume rainfall margins are heavy-tailed \cite[see e.g.][]{tencaliec19}, and within a region, we assume margins are asymptotically equivalent up to a constant. This last is a reasonable modeling assumption if we believe extreme episodes within a region have the same driver, let's say, a big storm.

\begin{figure}[htp]
\centering
\includegraphics[height=0.25\textwidth,width=1\textwidth]{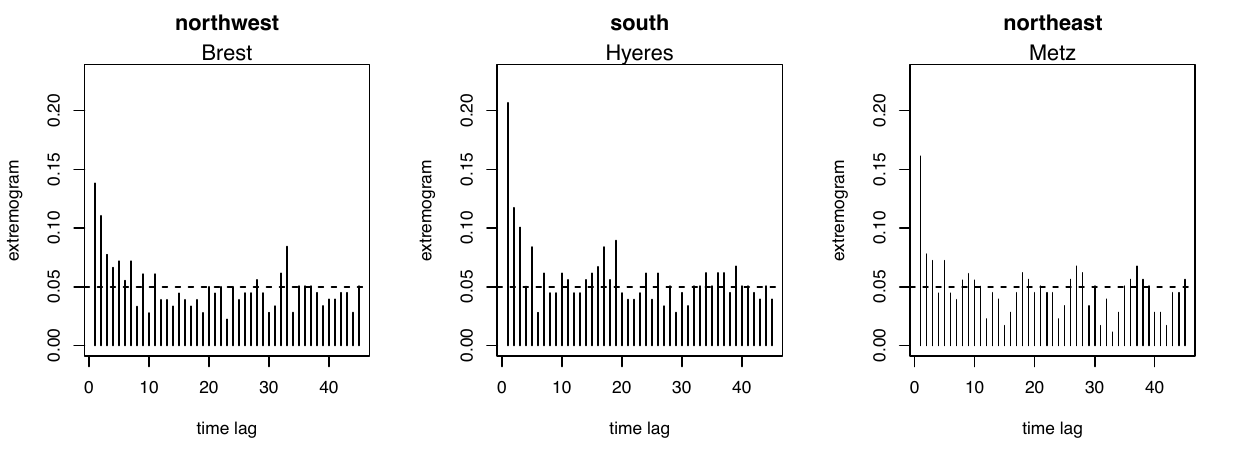}
   \caption{Empirical temporal extremogram of the $95$-th order statistic of fall daily rainfall in France from 1976 to 2015. The first, middle and last  correspond to three climatological regions as in Figure~\ref{fig:scatter_plot}. {As a baseline, the extremogram takes the value pointed by the dotted line on independent time lags.}
    }
   \label{fig:extremogram}
\end{figure}

{
Concerning the temporal ties, we see that at all nine stations, recording high rainfall levels at one day is often followed by measures from rainy days later since an extreme weather condition can last numerous hours. This extremal dependence in time is well illustrated by the temporal extremogram\footnote{The temporal extremogram is defined over time lags by $ t \mapsto \lim_{x \to + \infty}\P(X_t > x \, | \, X_0 > x)$.} introduced in \cite{davis:mikosch:2010} as can be seen in Figure~\ref{fig:extremogram}.}
Overall, we can explain the spatial and temporal links by the weather dynamics. As mentioned, it is reasonable to think that the main climatological event impacting an area often has the same source but is manifested at different time lags and locations. Our goal is to improve inference of its extremal features by a mindful aggregation of all measurements collected of it in space and time. We do so by introducing the stable sums method.

\par 

\section{Asymptotics of the stable sums method}\label{sub:sec:asymptotics}
In terms of notations, $(\bfX_t)_{t \in \mathbb{Z}}$ will always represent a stationary time series with tail index $\alpha > 0$, where $\bfX_t = (X_{t}(1), \cdots ,X_{t}(d))$ takes values in $\mathbb{R}^d$, that we endow with a norm $|\cdot|$. To model heavy-tailedness, we assume all vectors $(\bfX_t)_{|t| = 0,\dots, h}$ are multivariate regularly varying, $h\ge 0$; see Equation \eqref{eq:RV} for a precise definition.
For inference purposes, we consider the multivariate observations $(\bfX_1, \dots, \bfX_n)$, and we introduce a sum length sequence $(b_n)$, such that $n/b_n \to + \infty$, as $n \to + \infty$. For $p > 0$, we construct the sub-samples
\begin{align}\label{def:stableblock}
 \underbrace{\quad S_{1,b_n}(p)\quad}_{:= \sum_{t=1}^{b_n}|\bfX_t|^p} \;\;, \underbrace{\quad S_{2,b_n}(p)\quad}_{:=\sum_{t=b_n +1}^{2\,b_n}|\bfX_{t}|^p  } \;\;, \quad \cdots  \quad ,\underbrace{\quad S_{  \lfloor n/b_n \rfloor ,b_n}(p)\quad}_{:=\sum_{t= \lfloor n/b_n \rfloor-b_n+1}^{ \lfloor n/b_n \rfloor}|\bfX_{t}|^p }, 
\end{align}
with the convention $S_{b_n}(p):= S_{1,b_n}(p)$.
In this stationary regularly varying setting, \cite{buritica:mikosch:wintenberger:2021} proved that, for short-range memory time series, the following large deviations approximation holds
\begin{align}\label{eq:stablelimit}
\P( X_0(j) > x_{b_n})\;\approx\; m(j)\, (b_n \, c(p))^{-1}\, \P\big( S_{1,b_n}(p) > x_{b_n}^p  \big), 
\end{align}
as $n \to + \infty$, 
where $(x_n)$ corresponds to a suitable sequence satisfying $n\, \P(|\bfX_0| > x_n) \to 0$, as $n \to + \infty$, $m(j)$ takes values in $(0,1]$, for $j = 1, \dots, d$, and $p \mapsto c(p)$ is a decreasing function. Equation~\eqref{eq:stablelimit} models the tail of $X_0(j)$, the $j-$th coordinate of $\bfX_0$.

\par 
The practical key aspect of \eqref{eq:stablelimit} is that, whenever $m(j)$ and $c(p)$ are adequately estimated,  all extremal marginal features of the multivariate vector $\bfX_0$ can be easily deduced from the  single  univariate  sum $S_{1,b_n}(p)$.
For our case study, this means that any extreme   quantile of a  weather station, say  $j$,  can be directly deduced from  the sum $S_{1,b_n}(p)$ computed over  the group of three  neighbouring stations, 
albeit the knowledge 
of the two constants $m(j)$ and $c(p)$ in \eqref{eq:stablelimit}. 
We recall \cite{buritica:mikosch:wintenberger:2021} also showed $c(\alpha) = 1$, for short-range memory time series. Thus our goal is to motivate the choice $p=\alpha$ in \eqref{eq:stablelimit}. 
This modelling strategy obviously implies that the index of regular variation, $\alpha$, needs to be estimated as $\widehat{\alpha}^n$,
a necessary step in any Pareto-based quantile estimation. Then, albeit the knowledge of $m(j)$, the main challenge now is to infer the distribution 
\[ x \;\mapsto\; \P(  S_{1,b_n}(\widehat{\alpha}^n) \le  x), \] from the
the transformed  dataset $(S_{t,b_n}(\widehat{\alpha}^n))_{t=1,\dots, \lfloor n/b_n \rfloor}$. 
\par 
The natural question is then what is the appropriate model for $S_{b_n}(p)$, for $p > 0$. 
As  $S_{b_n}(p)$ is a sum of stationary regularly varying increments, then, assuming $n/b_n \to + \infty$, as $n \to + \infty$,  the  central limit theorem for weakly dependent stationary time series holds. 
There exists positive and real sequences $(a_n(p)), (d_n(p))$, such that 
\begin{align}\label{eq:stable:clt}
    S_{b_n}(p) - d_{b_n}(p))/a_{b_n}(p) \; \xrightarrow[]{d}\;  \xi_{\alpha/p}, 
\end{align}
where $\xi_{\alpha/p}$ is a stable distribution with stable parameter $\alpha/p$, the sequence $(a_n(p))$ satisfies $n\P(|\bfX_0|^p > a_{n}(p)) \to 1$, as $n \to + \infty$, and this convergence in distribution holds as the sums length $b_n$ goes to infinity.
Two important elements can be highlighted from this convergence.
First, the family of $\alpha$-stable distributions (see Section \ref{sec: stable}) appears as the natural parametric family to fit the sequence 
$(S_{t,b_n}(p))_{t=1,\dots, \lfloor n/b_n \rfloor}$. Second, the aforementioned choice of taking $p=\alpha$ is reinforced as the stable parameter $\alpha/p$ equals to one for this choice. 
This produces a solid  yardstick to select the right couple $\widehat{\alpha}^n,b_n$. In other words, an appropriate selection of $\widehat{\alpha}^n,b_n$, corresponds to the case 
when the distribution of 
$(S_{t,b_n}(\widehat{\alpha}^n))_{t=1,\dots, \lfloor n/b_n \rfloor}$ follows  a  stable  distribution with a stable unit parameter. 
The algorithm behind this strategy  will be explained in Section \ref{sec:parameter:tuning}. 
\par 
To interpret the two quantities $m(j), c(p)$, in Equation~\eqref{eq:stablelimit}, we write them as follows 
\begin{align}\label{eq:limit:clustering}
    \lim_{n \to + \infty}\frac{ \P( X_0(j) > x_n)}{\P(|\bfX_0| > x_n)} \;=\; m(j), \quad \lim_{n \to + \infty}\frac{\P((\, S_{1,n}(p)\,)^{1/p} > x_n)}{ n\, \P( |\bfX_0| > x_n)} \;=\; c(p),
\end{align}
such that $n \P(|\bfX_0| > x_n) \to 0$, as $n \to + \infty$.
{ 
The ratio between the norm feature $\P(|\bfX_0| > x_n)$ and the marginal feature $\P( X_0(j) > x_n)$ 
does not depend on $t$ (as   $t=0$), and consequently,
the constants $m(j)$ trace back the $d$-dimensional  structure of extremes, but not the temporal dynamic.}  
Recall for our case study, 
the three  stations within a region are assumed 
to have the same tail index and 
margins within the same region are assumed 
to be asymptotically equivalent, up to a constant. 
This  is in compliance with  the left-hand of \eqref{eq:limit:clustering}. 
Practically, this is justified by the close proximity among the three stations within each of our regions. 
Theoretically,  
the multivariate Breiman's theorem   \cite{fougeres:2012} tells us that tail equivalences can be obtained whenever a multiplicative or linear lighter-tailed noise impacts the variables at hand.
In contrast, the constant $c(p)$ captures, throughout  the $\ell^p$--norm,   the  temporal  clustering among extremes when compared to i.i.d. time series. Indeed, for all $p > 0$, if $(\bfX_t^\prime)_{t \in \mathbb{Z}}$ are i.i.d. distributed as $\bfX_0$ then the right-hand side of Equation~\eqref{eq:limit:clustering} equals one \cite[see e.g.][]{buritica:mikosch:wintenberger:2021}.
\par 
{
Notice that in our methodology, the choice of $p$ for $c(p)$ in Equation~\eqref{eq:limit:clustering} is up to the practitioner. The special case of $p=\infty$ is interpreted as taking the block of maxima, and has a strong connection with the so-called declustering techniques \cite[see e.g.][]{ferro:segers:2003}. Typically, the constant $c(\infty)$ equals the {\it extremal index} of the time series $(|\bfX_t|)_{t\in \mathbb{Z}}$, which has been understood as the reciprocate of the mean number of consecutive high levels recorded in a short period; cf. \cite{leadbetter:1983,leadbetter:lindgren:rootzen:1983}.
For univariate time series
the block of maxima can be modeled with classical extreme value theory based on generalized extreme value distributions \cite{coles:bawa:trenner:dorazio:2001}. However, this brings the difficult problem of inferring the extremal index.
On the other hand, to decluster the exceedances approach built on generalized Pareto distributions, applied studies typically base inference only on the maxima of clusters \cite{tendij:eastoe:tawn:randell:jonathan:2021}. However, if $c(\infty) < 1$ estimates of marginal features are biased \cite{eastoe:tawn:2012,fawcett:walshaw:2007,fawcett:walshaw:2012}. Instead, choosing $p=\alpha$ completely bypasses the estimation of $c(\infty)$ or any declustering strategy.
}

\par 
From a theoretical point of view, our method is motivated by equation \eqref{eq:stablelimit} proven in \cite{buritica:mikosch:wintenberger:2021}. 
We then use central limit theory to justify the parametric model for the partial sums. Borrowing classical telescopic sum arguments, we prove the limit with stable parameter one; see Theorem~\ref{thm:stable:limit},  which interests us as we take $p=\alpha$. This proof uses the {\it $\alpha$-cluster process} defined in \cite{buritica:mikosch:wintenberger:2021}. It simplifies the assumptions in \cite{bartkiewicz:jakubowski:mikosch:wintenberger:2011} and \cite{basrak:planinic:soulier:2018} who might have overlooked the unit stable domain, usually receiving less attention. Furthermore, \eqref{eq:stablelimit} justifies inference of extreme quantiles in the scope of the threshold sequence $(x_{n})$. Its order of magnitude was studied for classical examples as linear processes in \cite{mikosch:samorodnitsky:2000}, and for solutions to recurrence equations in \cite{buraczewski:damek:mikosch:zienkiewicz:2013,konstantinides:mikosch:2005}. 
For further references on large deviation probabilities for weakly dependent processes with no long-range dependence of extremes we refer to \cite{davis:hsing:1995,jakubowski:1997,jakubowski:1993,mikosch:wintenberger:2013}. Concerning central limit theory for stationary weakly dependent sequences, it was first addressed in \cite{davis:hsing:1995} using weak convergence of point processes.
Further, \cite{jakubowski:1997,jakubowski:1993} show central limit theory using classical telescopic sum arguments and large deviation limits. A modern treatment is conferred to \cite{bartkiewicz:jakubowski:mikosch:wintenberger:2011}. 

\section{Algorithm}\label{sec:stable:sums}
\subsection{Preliminaries}\label{sec: stable}
Let $\bfX 
$ be an $\mathbb{R}^d$-valued random vector. For $T > 0$, the multivariate $T$--return level is $\bfz_T = (z_T(1), \dots, z_T(d))$, where $z_T(j)$ is the $T$--return level associated to the $j$-th coordinate $z_T(j) = \inf\{z(j): \P( X(j) > z(j)) \le  1/T\}.$
We recall below the definition and basic properties of stable distributions.
\begin{definition}
The random variable  $\xi_a :=\xi_a (\sigma, \beta, \mu)$ 
follows 
a stable distribution with parameters $( \text{a} ,\sigma, \beta, \mu)$ if and only if, for all $u \in \mathbb{R}$,
\begin{align}\label{def:stable:characteristic:function}
    &\E\big[\exp\{ i \, u \, \xi_{a}\}\big] \nonumber\\
    &=
    \begin{cases}
     \exp \{ - \sigma^{a} |u|^{a} ( 1- i \, \beta \, \text{sign}(u) \, \tan \tfrac{\pi a }{2} ) + i \, \mu \, u \} &\text{if } a \not = 1, \\
     \exp \{ - \sigma |u|( 1- i \, \beta \, \text{sign}(u)\, \tfrac{2 }{\pi} \log |u| ) + i \,\mu\, u \} &\text{if } a  = 1,
    \end{cases} 
\end{align}
$a \in (0,2]$ is a stable parameter, $\sigma \in  [0,+\infty)$ is a scale parameter, $\beta \in [-1,1]$ is a skweness parameter,  and $\mu \in \mathbb{R}$ is a location parameter. 
\end{definition}
\par
Classical examples of stable distributions are the Gaussian distribution with $a=2$ and $\beta=0$, the Cauchy distribution with $a = 1$ and $\beta = 0$; and the L\'evy distribution with $a = 1/2$ and $\beta = 1$. 
Stable distributions satisfy the reflection property: if $ \xi_a := \xi_a(1,\beta,0)$ is a stable random variable with parameters $(a,1,\beta,0)$, then $-\xi_a$ is a stable random variables with parameters $(a,1,-\beta,0)$. 
The stable distribution is symmetric when $\beta = 0$, and has support in $\mathbb{R}$ when $|\beta| \not = 1$. If $\beta=1$ there are three cases: if $a < 1$ then the support of its density admits a finite lower bound. If $a = 1$ the density is supported in $\mathbb{R}$ but only the right tail is regularly varying. Otherwise, the stable distribution admits two heavy tails. A full summary on stable distributions can be found in \cite{feller:2008,nolan:2020,samorodnitsky:taqqu:linde:1996}.

\subsection{Model assumptions}\label{sec:model:assumptions}
{ In the remaining of the article we assume $(\bfX_t)_{t\in\mathbb{Z}}$ to be a stationary regularly varying time series taking values in} $(\mathbb{R}^d,|\cdot|)$, with index of regular variation $\alpha > 0$; cf. \cite{basrak:segers:2009}. This means there exists an $\mathbb{R}^d$-valued time series $(\bfTh_t)_{t \in \mathbb{Z}}$ such that $|\bfTh_0| = 1$ a.s., and
\begin{align}\label{eq:RV}
    \P((\bfX_t)_{|t| =0,\dots, h} \in \cdot \, | \, |\bfX_0| > x) \;\xrightarrow[]{d}\; \P(Y(\bfTh_t)_{|t| = 0, \dots, h} \in \cdot), \quad x \to + \infty,
\end{align}
where $Y$ is $(\alpha)$--Pareto distributed, $\P(Y > y) = y^{-{\alpha}}$, for all $y > 1$, independent of $(\bfTh_t)_{t \in \mathbb{Z}}$. We fix $|\cdot|$ to be the supremum norm, i.e. $|\bfX_0| := \max_{j=1,\dots,d}|X_0(j)|$, but any choice of norm is possible under minor modifications. We call $(\bfTh_t)_{t \in \mathbb{Z}}$ the spectral tail process.
\par 
For now, we suppose the approximation in \eqref{eq:stablelimit} holds and the renormalized process of partial sums $S_{b_n}(p)$ converges to a stable distribution with stable parameter $a=\alpha/p$, as $n \to + \infty$. These assumptions are satisfied for classical examples of weakly dependent regularly varying time series. We postpone the asymptotic theory behind it to Section~\ref{sec:Asymptotic:theory}.
Motivated by Theorem~\ref{thm:stable:limit}, we also set the  skweness parameter $\beta=1$ to simplify computations.


\subsection{Choice of the algorithm inputs}\label{sec:parameter:tuning}
To construct the sub-sample $(S_{t,b_n}(\alpha))_{t=1,\dots, \lfloor n/b_n \rfloor}$ defined in \eqref{def:stableblock}, for $p=\alpha$, we need to estimate the index of regular variation ${\alpha}$, and determine  the sum length $b_n$. Also, the indexes of spatial clustering $m(j)$ are required to use \eqref{eq:stablelimit}. 
\par 
We estimate $\alpha$ using the unbiased Hill-type estimator of  \cite{dehaan:mercadier:zhou:2016}, see their 
 Equation (4.2) of $\widehat{\alpha}^n$ that varies in function of order 
 statistics $k$. {Fixing the choice of $k$ we obtain a point estimate\footnote{Equation (4.2) in \cite{dehaan:mercadier:zhou:2016} yields to an estimate $\widehat{\alpha}^n(k)$, where $k$ is a fixed number of higher order statistics. We tune the second order parameter $\widehat{\rho}\le 0$ to the median value of $k_\rho \mapsto \widehat{\rho}(k_\rho)$, for $2 \le k_\rho \le k$; see \cite{gomes:dehaan:peng:2002,dehaan:mercadier:zhou:2016}. We then choose point estimate from a steady portion of the trajectory plot of $k \mapsto \widehat{\alpha}^n(k)$.} $\widehat{\alpha}^n = \widehat{\alpha}^n(k)$.}
To select the temporal window $b_n$,   
we recall the renormalized  partial sums, denoted
$(S_{t,b_n}(p))_{t=1,\dots, \lfloor n/b_n \rfloor}$, 
should follow,  for $p=\alpha$, a stable distribution with unit stable parameter, i.e., $a=1$. 
So, for a given $b_n$, we run a ratio likelihood test for the null hypothesis $(H_0):a=1$ and we only keep  pairs $\widehat{\alpha}^n, b_n$, such that the null hypothesis is not rejected at the $0.05$ level.
This heuristic allows us to discard an unsuitable choice for the couple $\widehat{\alpha}^n, b_n$. 

 Concerning the inference of $m(j)$, notice \eqref{eq:limit:clustering} and \eqref{eq:RV} yield  $ m(j) = \P(Y \Theta_0(j) > 1) =\E[(\Theta_0(j))^\alpha_{+}],$
 where $Y$ is $(\alpha)$-Pareto distributed, independent of the $d$--dimensional random variable $\bfTh_0$, and $|\bfTh_0|=1$ a.s. 
 In this context,   given   $\widehat{\alpha}^n$,   all $m(j)$, for $j=1,\dots,d$, are simply estimated by the following  empirical  means  
\begin{align}\label{em_estimates}
 \widehat{m}^n(j) := \frac{1}{k}\,\sum_{t=1}^n \, \frac{(X_t(j))^{\widehat{\alpha}^n}_{+} }{ |\bfX_t|^{\widehat{\alpha}^n}}\,
 \1(|\bfX_t| \,\ge\,  |\bfX_{(k)}|),
\end{align}
where  $|\bfX_{(k)}|$ is the $k$--th order statistic from the norm sample that we fix to be the $95$--th empirical quantile for the remaining of this article. For a review on inference of the spectral measure $\bfTh_0$, 
 we refer to \cite{buritica:mikosch:wintenberger:2021,davis:drees:segers:warchol:2018,drees:janssen:neblung:2021}.

\subsection{Algorithm}\label{sec: algo}
{We outline in  Algorithm~\ref{algo:RP2} the steps of the stable blocks method.}
\begin{algorithm}[ht!]\label{algo:RP2}
  \caption{Multivariate $T$--return level stable sums estimate} \label{al1}
  \KwIn{$(\bfX_1,\bfX_2,\dots,\bfX_n), b_n, \widehat{\alpha}^n$, $\widehat{ m}^n(1), \dots, \widehat{m}^n(d)$; see Sections \ref{sec:model:assumptions}, \ref{sec:parameter:tuning}\; } 
    compute $(S_{t,b_n}(\widehat{\alpha}^n))_{t=1,\dots, \lfloor n/b_n \rfloor }$ as in \eqref{def:stableblock} with $p = \widehat{\alpha}^n$,\\
     fit maximum likelihood stable parameters $\widehat{\theta}$,  $\widehat{\theta}^{a = 1}$,\\
     test the null hypothesis $(H_0): a = 1$ using ratio likelihood test,
     \uIf{  $(H_0)$ is not rejected: }{    
         $\widehat{\theta}= \widehat{\theta}^{ a = 1}$,\\ \label{eq:fitted:parameters}
         {
         \For{$j=1,\dots,d$\label{eq:5}  }
        {
         calculate $q_T(j)$ a $\widehat{\theta}$--stable quantile at $(1 - 1/(T\,\widehat{m}^n(j)) )^{b_n}$,\\       
         }
        }
        \Return $ \widehat{\bfz}^n_T := \big( (q_{T}(1))^{1/\widehat{\alpha}^n}, \dots, \, (q_{T}(d))^{1/\widehat{\alpha}^n}\big)$; see \eqref{eq:stablelimit}, \label{eq:10}
         
      }
      \Else{
      choose a different pair of parameters $\widehat{\alpha}^n, b_n$.
      }
 \end{algorithm} 

The multivariate $T$--return level is estimated applying Algorithm~\ref{algo:RP2} to $(\bfX_1,\bfX_2,\dots,\bfX_n)$. To obtain confidence intervals, we sample parametric bootstrap stable replicates with parameters $\widehat{\theta}$ as in line~\ref{eq:fitted:parameters}, and repeat the steps in lines \ref{eq:5} - \ref{eq:10} of
Algorithm~\ref{algo:RP2}. 
We use the percentile bootstrap method. 
A component-wise estimator is calculated applying Algorithm~\ref{algo:RP2} to $(X_1(j), \dots, X_n(j))$, for $j=1,\dots,d$.
Notice that for non-negative univariate time series, $m(1)=1$ from Equation~\eqref{eq:limit:clustering}, and both estimates coincide.
\par 
Concerning the asymptotic properties of the maximum likelihood estimator for stable distributed sequences, we refer to \cite{dumouchel:1971,dumouchel:1973a,dumouchel:1973b,dumouchel:1975} for large-sample theory. Bounds for the derivatives of the density function in terms of the parameters $(x;a,\sigma,\mu)$ have been computed therein; see also \cite{nolan:2001} for an overview on maximum likelihood methods for stable distributions.

\section{Simulation study}\label{sec:SS}
\subsection{Models}\label{subsec:models}
We consider the following models in our simulation.\\
    \noindent 
    {\bf Burr model:} Let $(X_1, \dots, X_n)$ be independent random variables distributed as $F$ with
    \begin{align}\label{eq:Burr:model}
      F(x ;  c , \kappa ) = 1 - \big( 1 + x^c  \big)^{-\kappa}, \quad x > 0,
    \end{align}    
    $ c,\kappa > 0$ are shape parameters thus $X_1$ is univariate regularly varying with index $\alpha = \tfrac{1}{c \kappa} > 0$.\\
    \noindent
    {\bf Fr\'echet model:} Let $(X_1, \dots, X_n)$ be independent random variables distributed as $F$ with
    \begin{align}\label{eq:Frechet:equation}
    F(x; {\alpha})  =e^{-x^{-{\alpha}}}, \quad x > 0,
    \end{align}
    then $X_1$ is univariate regularly varying with tail index ${\alpha} > 0$.\\
    \noindent
    {\bf ARMAX model:} Let  $(X_1, \dots, X_n)$ be sampled from the time series $(X_t)_{t \in \mathbb{Z}}$ defined as the stationary solution to the equation
    \begin{align}\label{ARMAX:equation}
    X_t = \max \big\{ \lambda\, X_{t-1},  \big(1-\lambda^{\alpha}\big)^{1/\alpha}\, Z_{t} \big\}, \quad t \in \mathbb{Z},
    \end{align}
    where $\lambda \in [0,1)$, and $(Z_t)_{t \in \mathbb{Z}}$ are independent identically distributed Fr\'echet innovations with tail index of regular variation $\alpha > 0$. Then $(X_t)_{t  \in \mathbb{Z} }$ is regularly varying with same index of regular variation but with extremal index equal to $1- \lambda^{\alpha}$. \\
    \noindent
    {\bf mARMAX$_\tau$ model:}  Let  $(\bfX_1, \dots, \bfX_n)$ be sampled from the time series $(\bfX_t)_{t \in \mathbb{Z}}$ defined as the stationary solution to the equation
    \begin{align}\label{def:marmax:lambda}
    X_t(j) := \max\big\{ \, \lambda(j)    X_{t-1}(j), \big(1-(\lambda(j))^{\alpha}\big)^{1/\alpha}\,  Z_t(j) \, \big\}, \qquad t \in \mathbb{Z},
    \end{align}
     for $j=1,\dots,d$, where $\lambda$ takes values in $[0,1)^{d}$ and $(\bfZ_t)_{t \in \mathbb{Z}}$ are independent identically distributed vectors from a Gumbel copula with Fr\'echet marginals and index of regular variation $\alpha >0$. Moreover, $\bfZ_1$ is distributed as $G$ defined by
    \begin{align}\label{def:marmax:tau}
    G(x; \alpha,\tau)= e^{-\big( (x(1))^{-(\alpha/\tau)} + (x(2))^{-(\alpha/\tau)} + \dots + (x(d))^{-(\alpha/\tau)}  \big)^\tau}, 
    \end{align}
     for $x \in \mathbb{R}^d$, and $\tau \in [0,1]$ that we refer as the coefficient of spatial dependence. The stationary solution $(\bfX_t)_{t \in \mathbb{Z}}$ is multivariate regularly varying with index of regular variation $\alpha > 0$; cf. \cite{ferreira:ferreira:2013} for more details.
     \par 
     Moreover, straightforward computations from \eqref{def:marmax:tau} yield
     \begin{align}\label{eq:tau}
     m(j)  = \lim_{x \to + \infty} \frac{\P(X_0(j) > x)}{\P(|\bfX_0| > x)} = \lim_{x \to + \infty} \frac{1-e^{-1/x^{\alpha}}}{1-e^{-d^{\, \tau}/x^{\alpha}}} = \frac{1}{d^{\,\tau}} < 1,
     \end{align}
     for all $j = 1,\dots,d$.
     {Then, from \eqref{eq:tau} we recover the symmetric properties of the Gumbel copula as $m(1)=\cdots = m(d) = 1/d^{\tau}$. We can also see from \eqref{eq:tau} that the coefficient of spatial dependence $\tau \in [0,1]$ plays a key role while measuring the spatial dependence of extremes. Indeed, similar calculations allow one to compute the spatial dependence parameter between any two marginals, say $j$, as 
     \begin{align}\label{eq:independence}
          \lim_{x \to + \infty} \P(X_0(j)>x\,|\,X_0(j^\prime)>x)= 2-2^\tau,
     \end{align}
     thus $\tau = 1$ points to asymptotic independence of extremes, whereas $\tau = 0$ indicates complete dependence of extremes. 
}

\subsection{Numerical experiment}\label{subsec:description}
{
  We perform a Monte Carlo simulation study with two main purposes. We aim to compare the stable sums method with the most common methods in applied studies based on declustering, see Section~\ref{sec:implementation:dec}. 
  We also aim to evaluate the stable sums multivariate approach compared to its component-wise implementation.
\par 
}
\par 
We estimate return levels $z_{T}$ for periods $T=20,50,100$ years of fall observations. This corresponds to the $99.95$-th, $99.98$-th and $99.99$-th quantiles. We simulate $1000$ trajectories of length $n = 4000$ from the models
presented in Section~\ref{subsec:models} with parameters:
\begin{itemize}
    \item Burr$(c,\kappa)$  with  $(c,\kappa)= (2,2)$ in \eqref{eq:Burr:model}.
    \item Fr\'echet$(\alpha)$ with ${\alpha} = 4$ in \eqref{eq:Frechet:equation}.
    \item  ARMAX$(\lambda)$ with $\alpha = 4$, for both $\lambda = 0.7$ and $\lambda = 0.8$ in \eqref{ARMAX:equation}.
    \item  mARMAX$_\tau(\lambda)$ taking values in $[0,+\infty)^3$ with $\alpha = 4$ and $\lambda = (0.7,0.7,0.7)$ in \eqref{def:marmax:lambda}, and for $\tau = 0.1,0.2,\dots,0.9$, in \eqref{def:marmax:tau}.
\end{itemize}
Notice ${\alpha} = 4$ in all the models above. This corresponds to a  typical rainfall tail index.
\par

\subsection{Implementation of stable sums method}\label{sec:implementation}
We fix the index of regular variation to be $\widehat{\alpha}^n = \widehat{\alpha}^n(k)$ (see Section~\ref{sec:parameter:tuning} for details) with $k = n^{0.7}$, for the Burr model, and $k = n^{0.9}$, for the Fr\'echet, ARMAX and mARMAX models.  Now notice that plugging in the estimates $\widehat{\alpha}^n, \widehat{m}^n$ in Algorithm~\ref{algo:RP2} we can run the stable sums method as a function of the sum lengths $b_n$. In this way, we implement our method for the sum lengths $bl = 2^i$, with $i = 4,5,6,7$. We sample $R = 100$ parametric bootstrap replicates to compute confidence intervals for the estimated return levels. For the multivariate models, we compute both the multivariate and component-wise stable sums estimator. 
  
 \subsection{Implementation of classical methods}\label{sec:implementation:dec}
{ For the univariate models, we also run the peaks over threshold and block maxima methods using the most popular declustering approaches \cite[see e.g.][]{coles:bawa:trenner:dorazio:2001}. 
A brief description of both implementation procedures is given below.
\par 
The peaks over threshold method models exceeded amounts over a high threshold with a generalized Pareto distribution; see chapters 4 and 5 in \cite{coles:bawa:trenner:dorazio:2001} for an overview. In our case, we fix the threshold level to be the $95$--th empirical quantile. For detecting clusters with various exceedances, we follow the ideas in \cite{ferro:segers:2003}. We keep only the largest peak from each cluster to correct confidence intervals, and fit a Pareto model to the exceeded amounts of this sub-sample. We use the code in the R-package {\it extRemes 2.0.12}, and our implementation follows the guide in \cite{gilleland:katz:2016}. We compute delta-method confidence intervals on the declustered sample. 
\par 
The block maxima method models the largest records form consecutive observations with a generalized extreme value distribution; see chapters 3 and 5 in \cite{coles:bawa:trenner:dorazio:2001} for an introduction. We implement it over disjoint blocks of length $bl_{BM}=20$. We estimate the extremal index using the interval's estimator in Ferro {et al.} \cite{ferro:segers:2003}, tuned with the $95$--th empirical quantile. We first fit a generalized extreme value distribution, then perform the extremal index estimation, and finally extrapolate high return levels using the R-package {\it extRemes 2.0.12}; see also the guide \cite{gilleland:katz:2016} for details. We compute delta-method confidence intervals.
}
\subsection{Simulation study in the univariate case}\label{results}
Estimation of the index of regular variation, as detailed in Section~\ref{sec:implementation}, yields unbiased estimates for the univariate models 
(plots can be available upon request). 
\par
\begin{figure}[!htb]
    \centering
    \includegraphics[height=0.58\textwidth,width=01\textwidth]{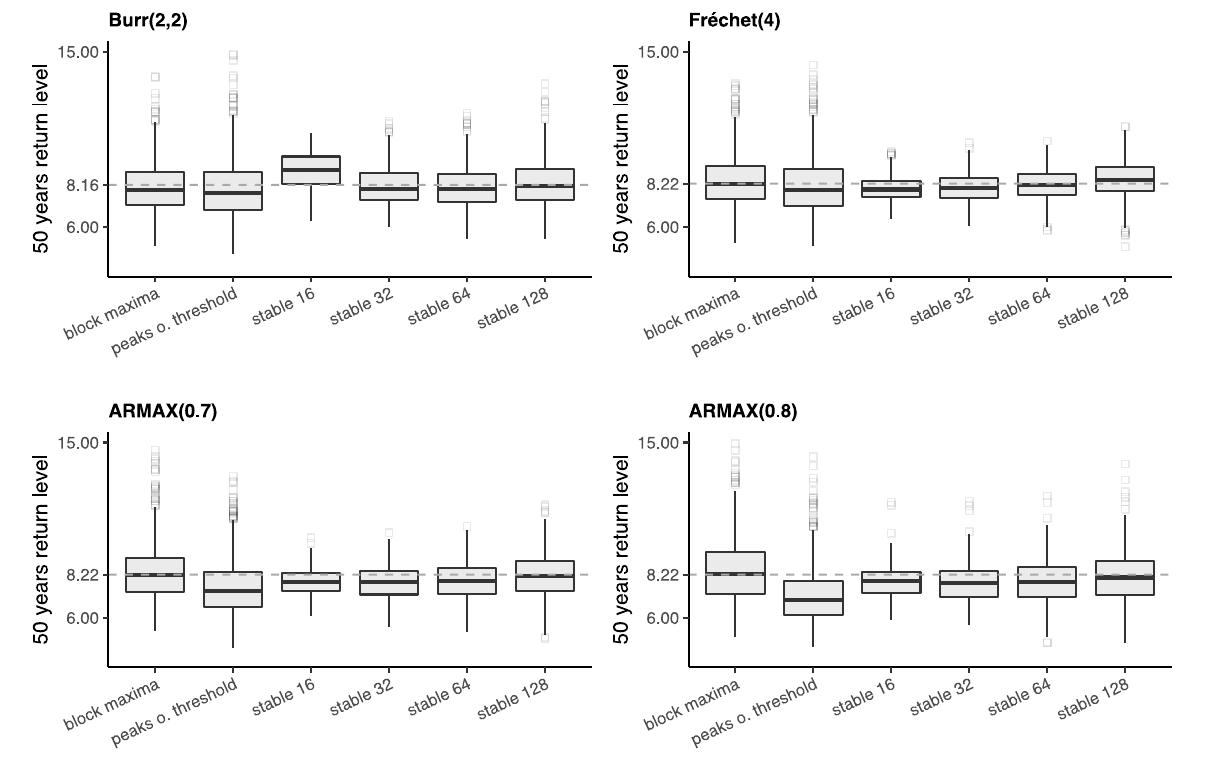}
         \caption{ Boxplots of estimates of $\widehat{z}^n_T$ with different methods where stable 16 refers to the stable sums method with sum length $b_l = 2^4=16$. Dotted lines indicate the true values.}
         \label{figstat}
\end{figure}

 
 We can see from Figure~\ref{figstat} that our method gives unbiased results and, as expected, the choice of the sum length can be seen as a trade-off between bias and variance. The median estimate of the 50 years return level with the peaks over threshold method underestimates the real value when implemented at the dependent models and this underrates the risk. {This bias was already observed in \cite{fawcett:walshaw:2012,fawcett:walshaw:2012}, which avert us from inferring marginal features from the maxima of clusters; see \cite{eastoe:tawn:2012}}. In comparison, our block maxima implementation gives satisfactory results for all four models regarding bias. However, it has a larger spread compared to the stable sums methods. We conclude that for all models Algorithm~\ref{algo:RP2} works fine coupled with a good estimate of the index of regular variation as the one detailed in Section~\ref{sec:parameter:tuning}.
\par 
\par
To evaluate the accuracy of confidence intervals from all methods, we compute the number of times they capture the correct value. One must keep in mind that for the stable sums method, Algorithm~\ref{algo:RP2} only returns an estimate if the test of the stable parameter equal to one is accepted. In this case, we also compute the proportion of acceptance of the ratio test among the $1000$ simulated trajectories from each model. We summarize the sample coverage probabilities in Table~\ref{tab:coverage}. The coverage results are not reliable when the proportion of test acceptance is small, however, it increases as the sum length increases. As a result, we notice from Table~\ref{tab:coverage} that we automatically discard the very small sum lengths. 
\par 
To sum up, we read from Table~\ref{tab:coverage} that the stable sums method outperforms the block maxima and peaks over threshold methods for sum lengths between $32$ and $64$, where acceptance of the ratio likelihood test is significant. Instead,  coverage probabilities are unsatisfactory for the peaks over threshold method, specially for the models with time dependence of extremes. The coverage for the block maxima method is not well calibrated and gives poor results for the Burr model which is the only model with a marginal distribution that does not belong to the family of generalized extreme value distributions. In this manner, we aim to point at the deficiency of the classical methods on small sample sizes. 
\begin{table}
\centering
\caption{Table of coverage probabilities. The value in parenthesis is the ratio test $(H_0):a=1$ acceptance proportion. In bold we highlight the optimal choice of sum length for the stable sums method. In our study, a precise coverage should be at $0.95$. \label{tab:coverage}}
\begin{tabular}{rrrrrrrrr}
 \multicolumn{2}{c}{Years}  & 20 & 50 & 100 &  & 20 & 50 & 100\\ \hline
& & \multicolumn{3}{c}{Burr(2,2)}  &   &\multicolumn{3}{c}{ Fr\'echet(4)} \\[2mm]
\; block maxima     &              & .91 & .89 & .87
         &              & .93 & .93 & .92\\
peaks o. threshold      &              & .87 & .85 & .83 
         &              & .89 & .87 & .86\\[5pt]
\qquad stable 16 &    $_{(.06)}$     & .89 & .85 & .80 
         &    $_{(.53)}$     & {.94} & {.95} & {.95}\\
\qquad stable 32 &   $_{(.51)}$     & {\bf.93} & {\bf.94} & {\bf.95}
         &    $_{(.83)}$     & {\bf.96} & {\bf.96} & {\bf.96} \\
\qquad stable 64 &   $_{(.85)}$     & {\bf.95} & {\bf.95} & {\bf.97} 
         &    $_{(.90)}$     & .96 & .99 & .99\\
\qquad stable 128&    $_{(.94)}$     & .87 & .98 & .98 
         &    $_{(.91)}$     & .82 & .99 & .99\\[10pt]  
& &  \multicolumn{3}{c}{Armax(0.7) }  & & \multicolumn{3}{c}{ Armax(0.8) }\\[2mm]      
\; block maxima      &              & .93 & .93 & .92
         &              & .92 & .91 & .91 \\
peaks o. threshold      &              & .78 & .79 & .79 
         &              & .66 & .72 & .74\\[5pt]
\qquad stable 16 &    $_{(.21)}$     & .92 & .94 & .93 
         &    $_{(.12)}$     & .80 & .82 & .84\\
\qquad stable 32 &    $_{(.66)}$     & .90 & .90 & .91
         &    $_{(.55)}$     & .87 & .89 & .90 \\
\qquad stable 64 &    $_{(.89)}$     & {\bf.93} & {\bf.96} & {\bf.96} 
         &    $_{(.85)}$     & {\bf.90} &{\bf .93} & {\bf.93}\\ 
\qquad stable 128&    $_{(.94)}$     & .85 & .97 & .98 
         &    $_{(.92)}$     & .83 & {.95} & {.96}\\[2mm]          
\end{tabular}
\end{table}

\subsection{Simulation study in the multivariate case}\label{results:multi}
\begin{figure}[!htb]
    \centering
    \includegraphics[height=0.34\textwidth,width=1\textwidth]{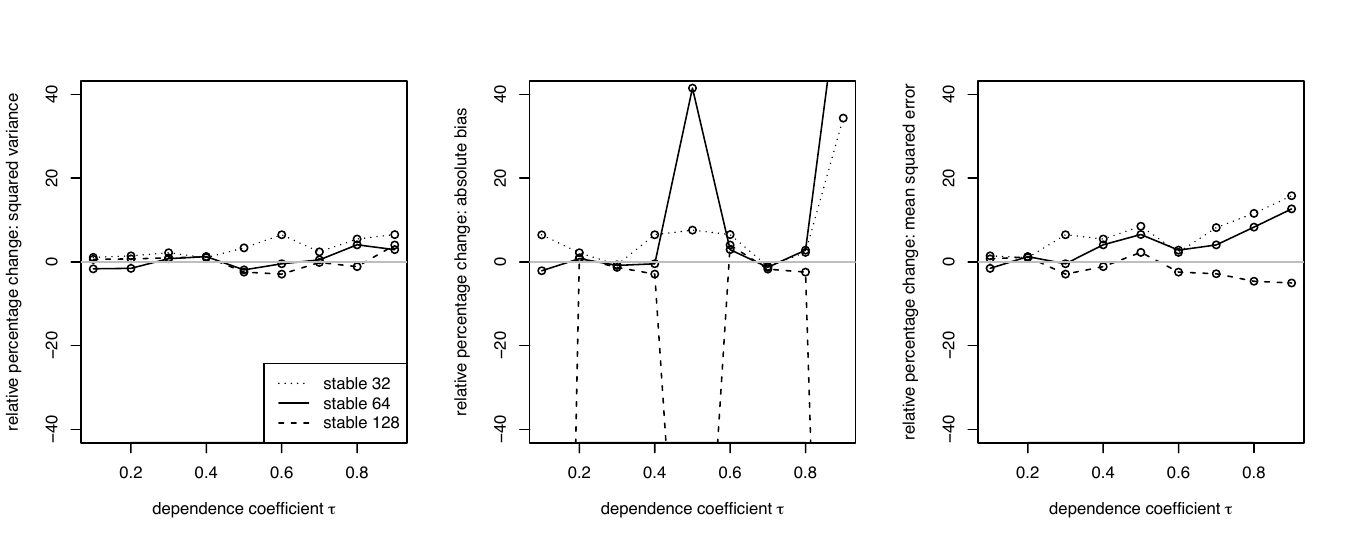}
         \caption{ Relative percentage change of the multivariate against the component-wise estimator (for the 50 years return level ${z}_{T}(3)$ estimator) of the: squared variance, absolute bias and mean squared error as in \eqref{eq:rel:perc:change}, from left to right. Positive values indicate a refinement while negative values indicate a degradation with the multivariate estimate.}
        \label{fig:relative:change}
\end{figure}

{
In this section, we aim to evaluate the pertinence of assessing extremal spatial dependencies. 
We inquiry now the performance of the multivariate, as opposed to  the component-wise, stable sums estimator.
We compute both estimates for the mARMAX$_\tau$ model samples with $\lambda = (0.7,0.7,0.7)$ as in \eqref{def:marmax:lambda}; see Section~\ref{subsec:description} for details. We compare the performance of both estimators at each coordinate, $j=1,2,3$, in terms of the mean squared errors relative percentage change. More precisely, for each coordinate, we compute mean squared errors of the multivariate and component-wise estimates denoted $MSE_{MV}$ and $MSE_{CW}$, respectively, and relate them by
\begin{align}\label{eq:rel:perc:change}
    \text{MSE relative percentage change} \;&= \; \frac{MSE_{CW} - MSE_{MV}}{MSE_{CW}} \times 100.
\end{align}
We also compute the relative percentage change of the squared variance, and of the absolute bias, from equations similar to \eqref{eq:rel:perc:change}. Large positive values point to an improvement of the multivariate estimator, while negative values detect a deterioration of its performance.
\par 
We omit details on coverage probabilities as they both have similar coverage as the ARMAX$(0.7)$ univariate model (as expected from  \eqref{def:marmax:lambda}). We analyze in detail estimates $z_T(3)$ of the $T=50$ years return level as similar results hold for all other coordinates. The relative percentage changes are plotted in figure~\ref{fig:relative:change} as a function of the spatial dependence coefficient $\tau$. We notice that for the sum lengths $32$ and $64$ the multivariate outperforms the component-wise estimator. Indeed, the choice of sum length $64$ was optimal for the ARMAX$(0.7)$ univariate model as pointed out by Table~\ref{tab:coverage}.
\par 
To conclude, for values of $\tau$ greater or equal than $0.4$, the multivariate outperforms the component-wise estimator for the optimal sum lengths of $32$ and $64$. As $\tau$ approaches $1$, and the model approaches the regime of asymptotic independence, the multivariate estimator has an outstanding improvement. If $\tau = 1$, the assumption of asymptotic independence means that we work with $3$ independent time series, identically distributed as $(X_t(3))$, which are sampled from the ARMAX$(0.7)$ model. For this reason, it is reasonable to obtain a gain from aggregating spatial extremes. 
In contrast, the amelioration is less evident for values of $\tau$ close to $0$. Recall from equation \eqref{eq:independence} that $\tau = 0$ points to asymptotic dependence.
{
We conclude that in general the multivariate estimator is preferable to the component-wise approach since it also has a gain in computational time.
Identifying, both theoretically and practically, which spatial features efficiently improve the multivariate inference procedures requires further investigation.
}
}

\section{Case study of heavy rainfall in France}\label{sec:case:study}
{ We recall the data set of fall daily rainfall introduced in Section~\ref{sec:introduction} and our goal of computing the level of daily rain to be exceeded in 50 years at all the nine weather stations in France. We conduct our analysis separately over the three different regions: northwest, south, and northeast of France. 
Fall observations from the same region are modelled as a 3-dimensional sample $(\bfX_1, \dots, \bfX_n)$, from a stationary multivariate regularly varying time series, i.e., $\bfX_t := (X_t(1), X_t(2), X_t(3))$, $t \in \mathbb{Z}$. 
We include both wet and dry days in our daily observations. 
In this setting, our goal 
traduces to estimating the $99.98$-th quantile of $X_0(j)$, for $j=1,2,3$.
}
\subsection{Implementation}\label{sec:tuning:par}
\par
 To study the $3$-dimensional sample obtained from each region $(\bfX_1,\dots,\bfX_n)$, we implement the stable sums method as a function of $k$ in the following way. For $k = 150, 250, 350, 450, 550$, first, we compute estimates $\widehat{\alpha}^n(k)$ as described in Section~\ref{sec:parameter:tuning}, second, we search the sum length larger than $32$ for which the $p-$value of the ratio likelihood test from Algorithm~\ref{algo:RP2} is minimized, among the first 20 acceptances of the test. We obtain in this way couples $\widehat{\alpha}^n(k), b(k)$. 
\par 
\par 

\subsection{Analysis of the radial component}\label{sec:radial}
\begin{figure}[!htb]
    \centering
    \includegraphics[height=1\textwidth,width=0.45\textwidth,trim={14cm 0 0 0},clip]{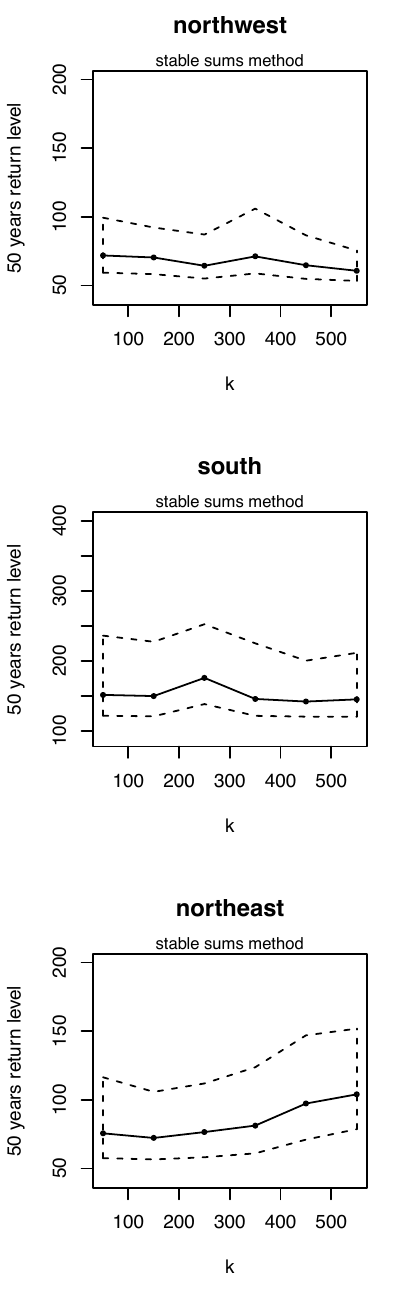}
   \caption{ 
   Estimates of the 50 years return level of fall supremum norm observations with confidence intervals. We write estimates as a function of $k$ with the parametrization detailed in Section~\ref{sec:tuning:par}.  }
   \label{fig:resultCI}
\end{figure}

At each region, we start by studying the supremum norm sample $(|\bfX_1|, \dots, |\bfX_n|)$. The estimates of the 50 years norm return levels are presented in Figure~\ref{fig:resultCI} as a function of $k$. The rows correspond to different regions.
We see that the stable sums method gives robust estimates as a function of $k$.
For comparison, the Pareto-based methods are also implemented in terms of $k$ in the supplementary material.
\par
\begin{figure}[!htb]
    \centering
    \includegraphics[width=0.9\textwidth, height=0.63\textwidth]{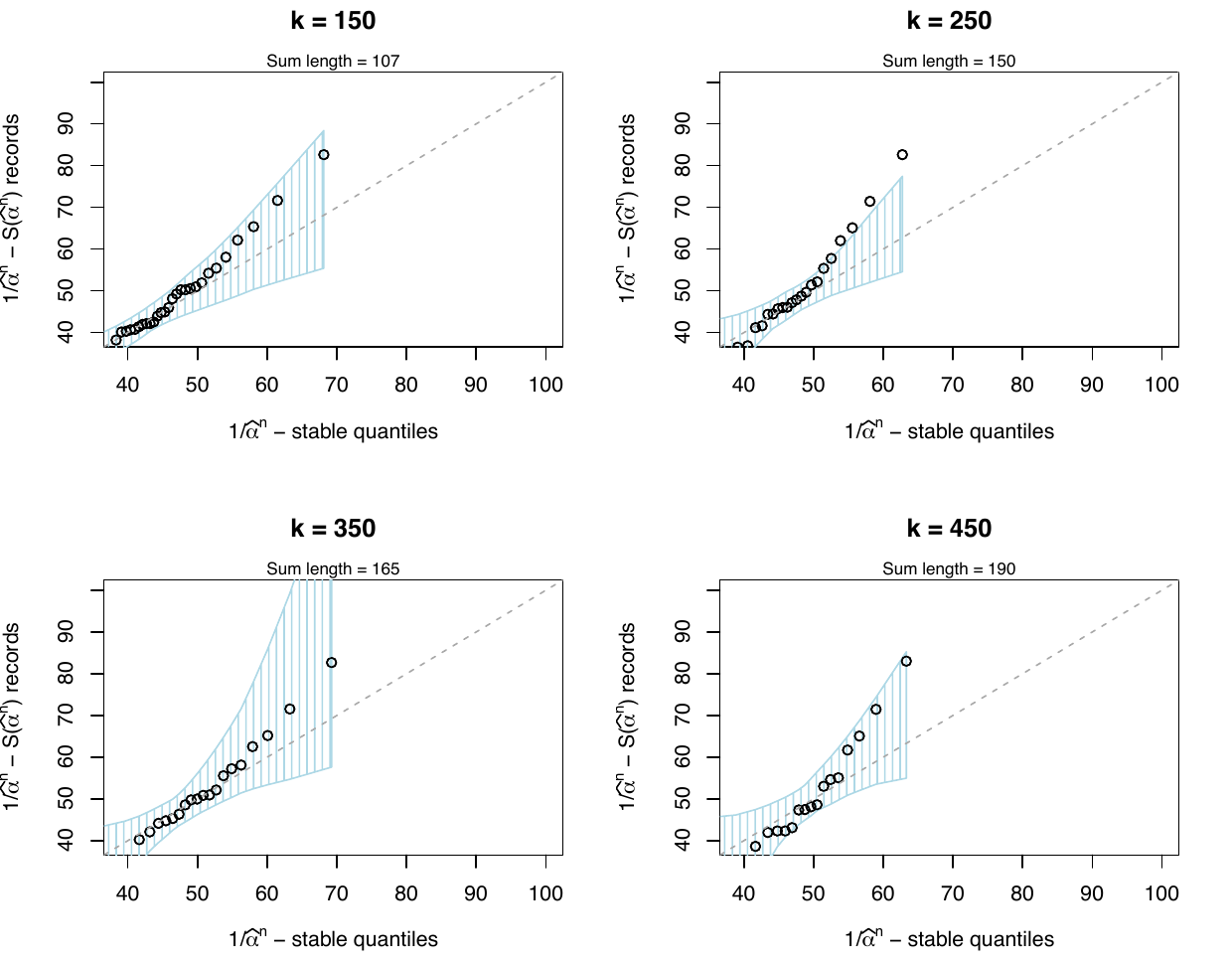}
   \caption{
   qqplots for different $k$ values of the $1/\widehat{\alpha}^n(k)$--stable quantiles against the $1/\widehat{\alpha}^n(k)$--$(S_{t,b}(\widehat{\alpha}^n(k)))_{t=1,\dots,\lfloor n/b\rfloor}$ records with $95\%$ confidence intervals for the northwest.}
   \label{fig:resultQQ1}
\end{figure}

\begin{figure}[!htb]
    \centering
    \includegraphics[width=0.9\textwidth, height=0.63\textwidth]{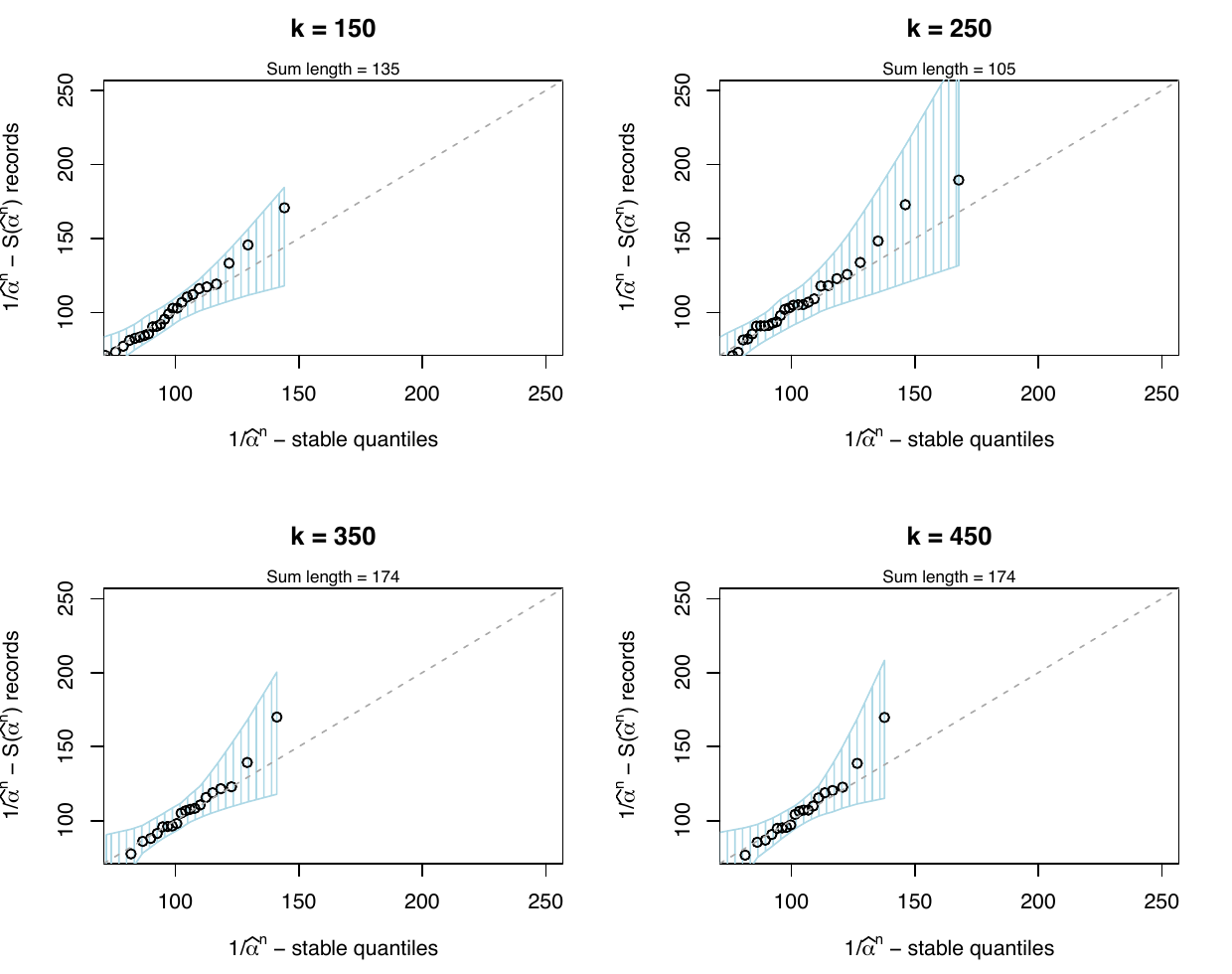}
   \caption{
   qqplots for different $k$ values as in Figure~\ref{fig:resultQQ1} but for the southern region.}
   \label{fig:resultQQ2}
\end{figure}

\begin{figure}[!htb]
    \centering
    \includegraphics[width=0.9\textwidth, height=0.63\textwidth]{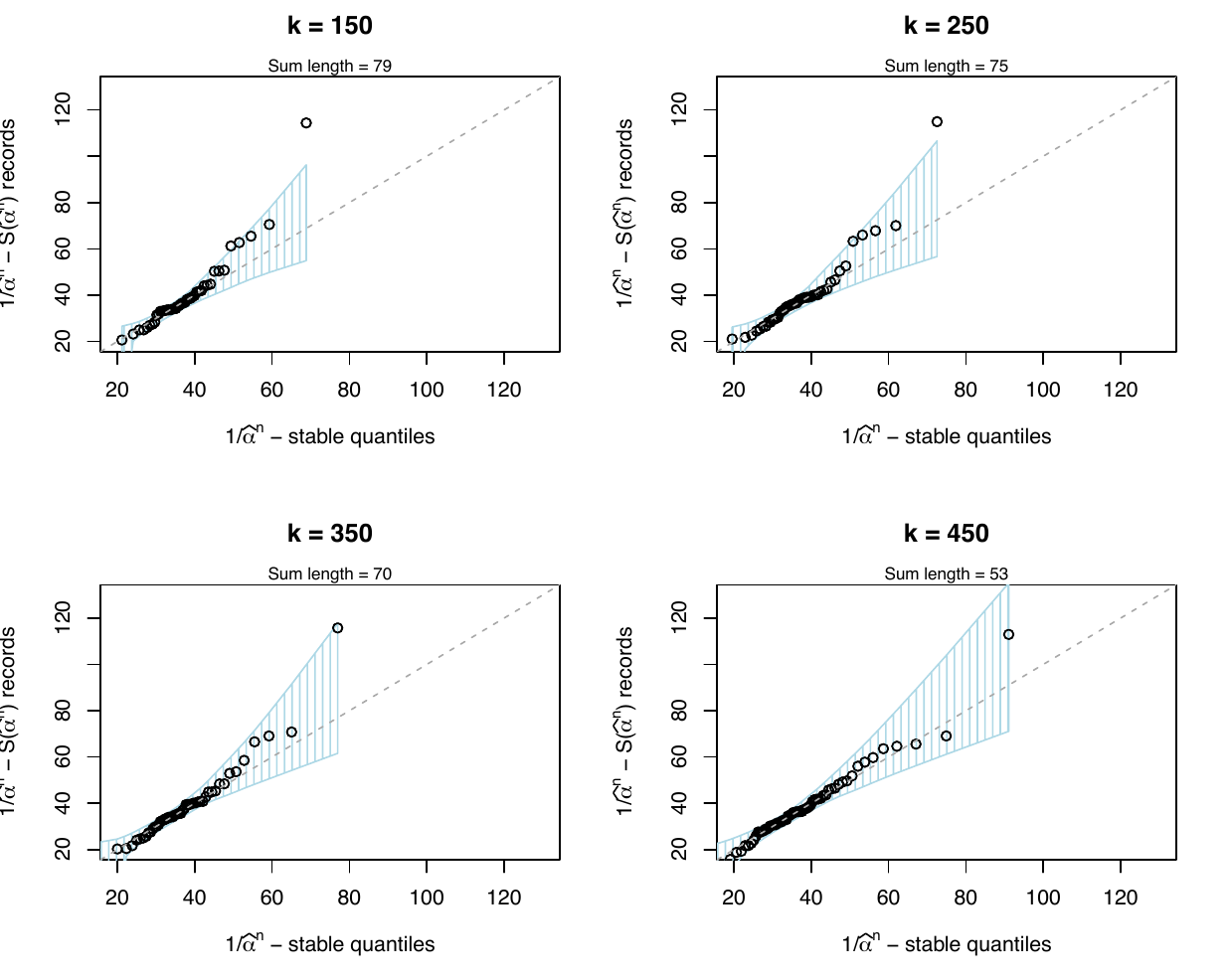}
   \caption{ 
   qqplots for different $k$ values as as in Figure~\ref{fig:resultQQ1}  but for the northeast region.}
   \label{fig:resultQQ3}
\end{figure}

 
To select a value for $k$, we inspect the qqplots of the observed stable records given by $(S_{i,b(k)}(\widehat{\alpha}^n(k))^{1/\widehat{\alpha}^n(k)})_{t=1,\dots,\lfloor n/b(k) \rfloor }$, against the theoretical stable quantiles to the power $1/\widehat{\alpha}^n(k)$. Recall the pairs $\widehat{\alpha}^n(k), b(k)$ are the ones detailed in Section~\ref{sec:tuning:par}. Figures \ref{fig:resultQQ1}, \ref{fig:resultQQ2}, \ref{fig:resultQQ3} contain the qqplots for the northwest, south and northeast, respectively, and allow us to assess goodness of fit for the different choices of $k$. We conclude from Figure~\ref{fig:resultQQ1} that for the northwest locations, the choice $k=350$, $b(k) = 165$ captures nicely the intermediate and extreme quantiles. For the southern region, we see in Figure~\ref{fig:resultQQ2} that the choice $k = 150$ and $b(k) = 135$ gives an accurate fit. Lastly, for the northeast region, Figure~\ref{fig:resultQQ3} suggests the choices $k=350$ and $b(k) = 70$, or $k=450$ and $b(k)=53$, for a correct alignment of intermediate and high quantiles.
\par 



\subsection{Analysis of the multivariate components}
\begin{figure}[!htb]
    \centering
   \includegraphics[height=0.33\textwidth,width=.9\textwidth]{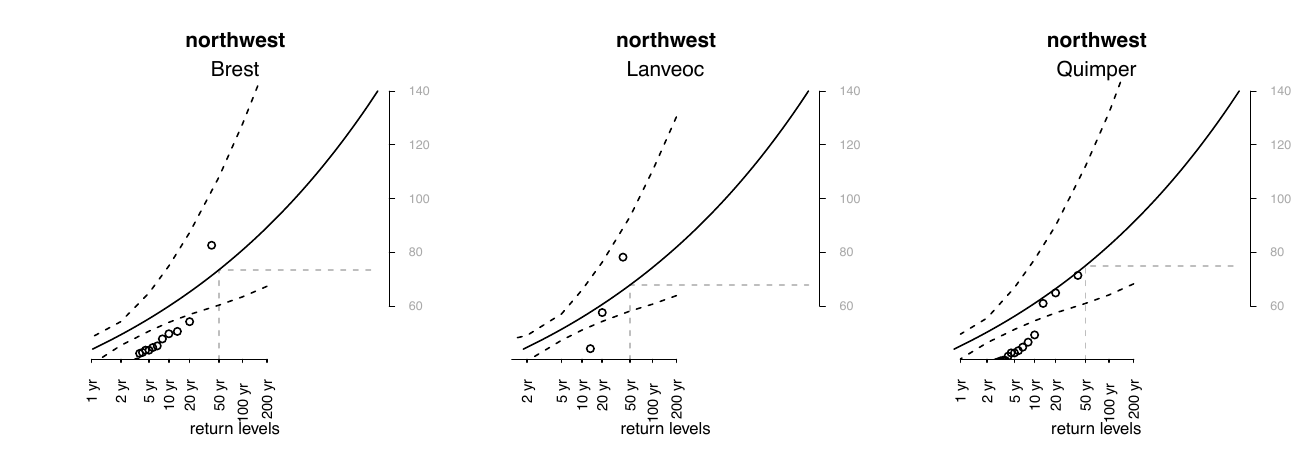}\\[\smallskipamount]
    \includegraphics[height=0.33\textwidth,width=.9\textwidth]{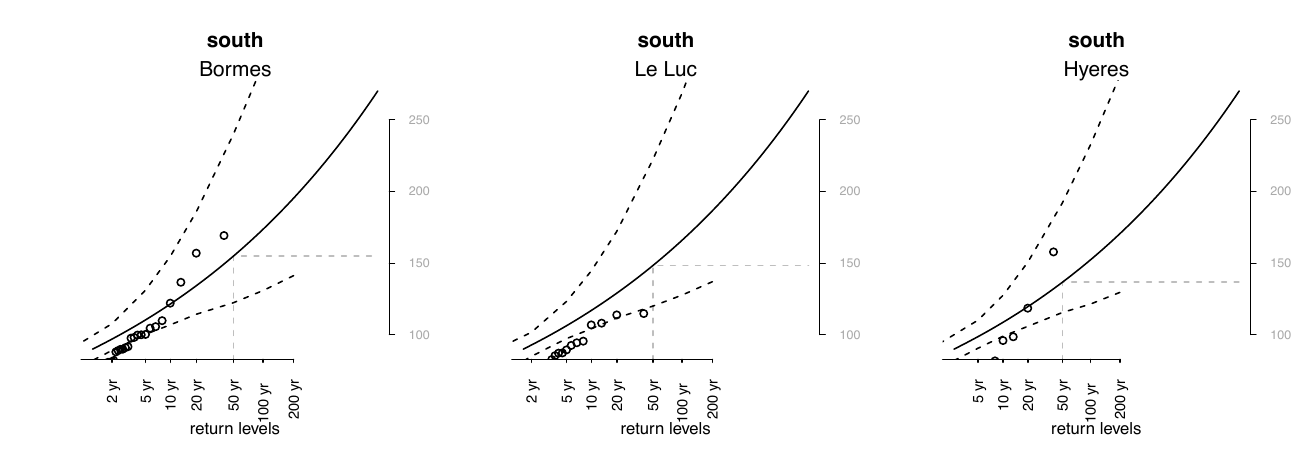}\\[\smallskipamount]
     \includegraphics[height=0.33\textwidth,width=.9\textwidth]{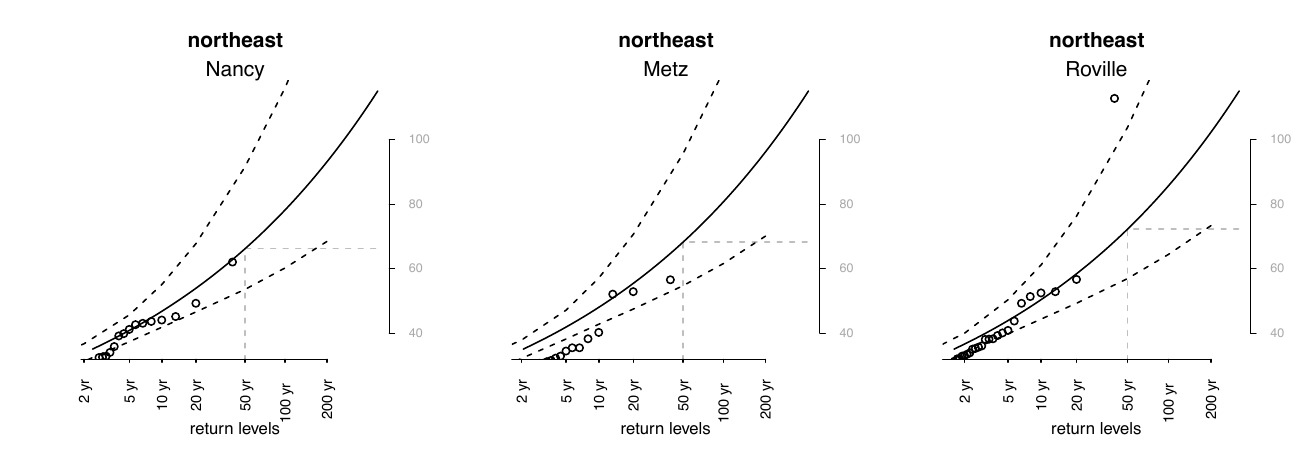}\\[\smallskipamount]
   \caption{
   Return level plots with confidence intervals  based on $1\,000$ bootstrap replicates. The grey dotted line gives the estimated 50 years return level.  Other return levels can be obtained alike.  The black line is the $\log(x) \mapsto x $ plot. The points are the logarithm of estimated  $1/\widehat{\alpha}^n$ - stable quantile at $(1- \tfrac{k}{n\,\widehat{m}^n(j)})^{b_n}$ vs. the $k$-th largest order statistics; see Equation~\eqref{eq:quantiles:2}. The largest order statistics should lie on the solid line. }
   \label{fig:multi}
\end{figure}

Finally, we turn back to
the component-wise analysis. In this case, we must estimate the indexes of spatial clustering. Relying on \eqref{em_estimates}, we obtain estimates: $\widehat{m}^n = (0.4966, 0.2709, 0.5744)$ corresponding to the weather stations at Brest, Lanveoc and Quimper in the northwest region; $\widehat{m}^n = (0.6064, 0.4706, 0.2866)$ for Bormes, Le Luc and Hyeres in the south; and $\widehat{m}^n = (0.3910, 0.4448, 0.5649)$ for Nancy, Metz, Roville in the northeast. Figure~\ref{fig:multi} plots the estimated return levels and confidence intervals for each station based on estimates $\widehat{m}^n$, and the tuning parameters $\widehat{\alpha}^n, b(k)$ from Section~\ref{sec:radial}, pointing to a nice fit of the radial component. In particular, we fix $k=350$, $b(k)= 165$ for the northwest region, $k=150$, $b(k) = 135$ for the south and $k=350$, $b(k)=70$ for the northeast.  Rows correspond to different regions. 
\par 
Moreover, we interpret \eqref{eq:stablelimit}, roughly speaking, to say: heavy daily rainfall at each weather station can be modelled as high quantiles of a stable distribution. In particular, letting the largest order statistics from each station play the role of the sequence of high threshold levels in \eqref{eq:stablelimit}, we deduce the following empirical version of this relation
\begin{align}\label{eq:quantiles}
    \P\big( (  S_{b_n}(\widehat{\alpha}^n) )^{1/\widehat{\alpha}^n } \le  X_{(k)}(j) \big) \; &\approx \;  1- \tfrac{k}{ \widehat{m}^n(j) \, n/b_n}\\  \label{eq:quantiles:2} 
    &\approx \big(  1 - \tfrac{k}{ \widehat{m}^n(j)\, n}\big)^{b_n}
\end{align}
 where $X_{(1)}(j) \ge X_{(2)}(j) \ge \cdots \ge  X_{(n)}(j)$, for $j = 1,\dots,d$, and \eqref{eq:quantiles:2} holds whenever $n/k_n \to + \infty$, and $b_n/n \to + \infty$, as $n \to + \infty$. Indeed, the approximations in \eqref{eq:quantiles}, \eqref{eq:quantiles:2} are only justified for large observations; see Section \ref{sec:model:assumptions}. The left-hand side in \eqref{eq:quantiles} can be approximated using the stable fit in line~\ref{eq:fitted:parameters} of Algorithm~\ref{algo:RP2}. In this case, \eqref{eq:quantiles:2} yields the estimated $T$-return level $j$-th coordinate from Algorithm~\ref{algo:RP2}, with $T=n/k$. We work using \eqref{eq:quantiles:2} to have a positive difference $1-1/(\widehat{m}^nT)$ in a wide range. We inspect \eqref{eq:quantiles:2} in Figure~\ref{fig:multi} by adding the $1/\widehat{\alpha}^n$--stable quantiles, from the fit in line~\ref{eq:fitted:parameters} of Algorithm~\ref{algo:RP2}, against the sample largest order statistics. 
We interpret the largest records close to the solid line as a nice fit. 
\par 
Overall, the most extreme records lie inside the confidence bands. The intermediate quantiles shouldn't necessarily align, and in practice, there is not a clear procedure for knowing how many of the top quantiles should line up with the solid line in Figure~\ref{fig:multi}.  
We conclude that in generate the multivariate method captures accurately the highest rainfall records, and supported by the numerical results from Section~\ref{results:multi}, it is justified for addressing the spatio-temporal dependencies of extremes.

\section{Asymptotic theory}\label{sec:Asymptotic:theory}
In the remaining, we discuss the theory behind the stable sums method. We will denote $(\bfX_t)_{t \in \mathbb{Z}}$ to be a regularly varying time series in $(\mathbb{R}^d,|\cdot|)$ as in \eqref{eq:RV}.
\par 
Theorem~\ref{thm:stable:limit} below states that under classical conditions \cite[see e.g.][]{davis:hsing:1995,bartkiewicz:jakubowski:mikosch:wintenberger:2011}, the asymptotics assumed in Section~\ref{sec:model:assumptions} hold for $p=\alpha$ (for $p \not = \alpha$ see \cite{buritica:mikosch:wintenberger:2021}). We introduce the anti-clustering, vanishing-small-values and mixing conditions: ${\bf AC}, {\bf CS}, {\bf MX}$, respectively, and comment on them below. A full discussion, and the proof of Theorem~\ref{thm:stable:limit} can be found in the supplementary material.  

\begin{theorem}\label{thm:stable:limit}
Let $(\bfX_t)_{t \in \mathbb{Z}}$ be a regularly varying time series in $(\mathbb{R}^d,|\cdot|)$ with (tail)-index $\alpha > 0$. Let $Z_t = |\mathbf X|^\alpha, t \in \mathbb{Z}$, and let $(a_n(\alpha))$ be such that $n\P(Z_t > a_n(\alpha)) \to 1$. Assume there exists $(k_n)$ satisfying $n/k_n \to +\infty$, as $n \to + \infty$, and assume ${\bf AC} (a_n(\alpha)), {\bf CS}(a_n(\alpha)), {\bf MX}(k_n)$ hold, i.e., for all $\epsilon, \delta  > 0$,  $u \in \mathbb{R}$,\\[2pt]
\noindent
{\bf AC} $(a_n)$:
$\lim_{l \to \infty}\limsup_{n \to \infty}\P(\max_{t=l,\dots,k_n } |Z_t|  > \epsilon  a_{n} | |Z_0| > \epsilon  a_{n} ) = 0. $\\
\noindent
${\bf CS}$\, $(a_n)$:
    $ \lim_{\epsilon \downarrow 0} \limsup_{n \to \infty} \frac{  \P( \sum_{t=1}^{k_n} |Z_t| \1_{\{|Z_t| \le \epsilon  \, a_{n}\}}  >\delta \, a_{n} )}{k_n\, \P( |Z_t| >  a_{n}) } \;=\; 0. $\\
\noindent    
${\bf MX} (k_n)$: $ \lim_{n \to \infty} |\E [e^{ i \, u  \sum_{t=1}^{n}  Z_t/a_n } ]  - \E [ e^{ i \, u  \sum_{t=1}^{k_n}  Z_t/a_n } ]^{\lfloor n/k_n \rfloor}| \;=\; 0\,$. \\[2pt]
\noindent
thus $k_n\P(Z_t > a_n(\alpha)) \to 0$, as $n \to + \infty$, and
\[
(S_{1,n}({\alpha}) - d_n(\alpha))/a_n(\alpha) \;\xrightarrow[]{d}\;\xi_1, \quad n \to + \infty.
\]
where $d_n(\alpha) = \E[Z_t\1( Z_t \le a_n(\alpha) )]$, and $\xi_1$ is stable distributed with stable and skewness parameters $a = 1$ and $\beta=1$; see \eqref{def:stable:characteristic:function}. Moreover, assuming {$\bf AC$}$((x_{k_n})^{\alpha})$, {$\bf CS$}$((x_{k_n})^{\alpha})$ we deduce \eqref{eq:stablelimit} holds for levels $(x_n)$ satisfying $n\P(|\bfX_0| > x_n) \to 0$, as $n \to + \infty$.
\end{theorem}

\begin{remark}
Condition ${\bf AC}$ is tailored to avoid long-range extremal dependence \cite[see e.g.][]{basrak:segers:2009,buritica:meyer:mikosch:wintenberger:2021}. 
It is also a common assumption to justify the declustering procedures from Section~\ref{sec:implementation:dec}; see \cite{ferro:segers:2003}. It holds for short-range memory series. Consider $m_0$-dependent stationary regularly varying sequences $(Z_t)_{t \in \mathbb{Z}}$, for example, the moving average
\begin{align}\label{eq:m0:dependence}
 Z_t = \varphi_0 Z_{t}^\prime + \cdots + \varphi_{m_0} Z^\prime_{t-m_0}, \quad t \in \mathbb{Z},
\end{align}
where  $(Z^\prime_t)_{t \in \mathbb{Z}}$ is an i.i.d. sequence distributed as a heavy-tailed random variable $Z_0$. In this case, condition ${\bf AC}(x_n)$ holds for any $(x_n)$ such that $n\P(|Z_t| > x_n) \to 0$, as $n \to + \infty$.
\end{remark}
\begin{remark}
Condition ${\bf CS}$ help us dealing with the asymptotics of sums of regularly varying sequences. Similar conditions were also considered in \cite{bartkiewicz:jakubowski:mikosch:wintenberger:2011, davis:hsing:1995, mikosch:wintenberger:2013,buritica:mikosch:wintenberger:2021,buraczewski:damek:mikosch:zienkiewicz:2013}. For $m_0$-dependent regularly varying time series  $(Z_t)_{t \in \mathbb{Z}}$ (see e.g. \eqref{eq:m0:dependence}), condition ${\bf CS}(x_n)$ holds for sequences $(x_n)$ such that $n\E[|Z_0/x_n|\1_{\{Z_0 \le x_n\}}] \to 0$, which implies $S_n(\alpha)/x_n \xrightarrow[]{\mathbb{P}} 0$, as $n \to + \infty$. This follows from Remark 5.2. in \cite{buritica:mikosch:wintenberger:2021}. In particular, the limit expectation equals zero if there exists $\kappa > 0$ such that $n/x_n^{1-\kappa} \to 0$. 
\end{remark}

\begin{remark}
     In our regularly varying setting, condition ${\bf MX}$ is common to many proofs of central limit theory (see condition (2.8) in \cite{bartkiewicz:jakubowski:mikosch:wintenberger:2011}). Actually, it has been verified on numerous examples under mixing-type assumptions; cf. \cite{bartkiewicz:jakubowski:mikosch:wintenberger:2011,mikosch:wintenberger:2013} and references therein. 
     In particular, it holds whenever the decay of the mixing coefficients\footnote{ The mixing coefficients $(\alpha_h)$ are defined, for all $h \in \mathbb{N}$, as
    \[
     \alpha_h \;:=\; \sup_{A \in \sigma( \,(X_t)_{t \le 0}\,), B \in \sigma(\, (X_t)_{t \ge h }\,)} |\P(A\cap B) - \P(A)\P(B)|.
     \]} $(\alpha_h)$ happens sufficiently fast; cf. Lemma 3.8. in \cite{bartkiewicz:jakubowski:mikosch:wintenberger:2011}. For $m_0$-dependent regularly varying time series $(Z_t)_{t \in \mathbb{Z}}$ (see e.g. \eqref{eq:m0:dependence}), it is easy to see that condition ${\bf MX}(k_n)$ holds choosing $k_n > m_0$.
 
\end{remark}

\section{Conclusions}\label{sec:conclusion}
{
Atmospheric conditions drive the heavy-rainfall measurements. These records have a spatial and temporal coverage explained by the storm/fonts dynamics. Typically, an extreme event with a common source is recorded simultaneously at different locations and over different time lags. 
In this work, we have proposed the stable sums method to aggregate space and time information of dependent observations. Our ultimate goal was to extrapolate high quantiles at each weather station. 
\par 
Our approach relies on the asymptotics of $\alpha$-power sums of regularly varying increments (i.e., we let $p=\alpha$ in \eqref{eq:stablelimit}). A parametric model for the sums $(S_{t,b_n}(\alpha))$ is at hand thanks to Theorem~\ref{thm:stable:limit}. Our method has proven to be robust for dealing with time dependencies. For statistical applications based on time-dependent observations, our method has made integrating the multidimensional aspects of extremes manageable. 
\par 
Our stable sums approach could also  be used to address other environmental extremal  problems. We now comment on one of them based on the idea that  
the asymptotics of space and time multivariate extreme events can be  summarized  by the univariate random variable of partial sums. 
In this work, we allocated  weights $m(j)$ to each coordinate to compute the marginal features 
like the set  $\{X(j) > x\}$.
Conceptually, it should be  also possible to  study other $d$-dimensional extremal sets, for example, $\{X(j) > x, X(j\prime) > x \}$. Applying the theoretical  results of \cite{buritica:mikosch:wintenberger:2021} will introduce  weights of the type $m(j,j^\prime)$. Still, our  take-home message will remain the same:    important  $d-$dimensional features are accessible by fitting only the univariate sums. 
}

\bibliographystyle{plain}
\bibliography{paper-ref}

\end{document}